\documentclass[aps,prb,showpacs,amsmath,twocolumn,amssymb,superscriptaddress,letterpaper]{revtex4}
 \usepackage{mathrsfs}
 \usepackage{graphicx}
 \usepackage{graphics}
 \usepackage{subfigure}
 \usepackage{bm}
 \usepackage{dcolumn}
 \usepackage{amsmath,bm}
 \usepackage{color}
 \usepackage{color}

\begin{document}
 	\title{Quantum to classical crossover under dephasing effects in a two-dimensional percolation model }
 	
\author{Junjie Qi}
\affiliation{International Center for Quantum Materials and School of Physics, Peking University, Beijing 100871, China}
\affiliation{ Collaborative Innovation Center of Quantum Matter, Beijing, China}
\author{Haiwen Liu}
\affiliation{Center for Advanced Quantum Studies, Department of Physics, Beijing Normal University, Beijing, 100875, China}
\author{Chui-zhen Chen}
\affiliation{Institute for Advanced Study and School of Physical Science and Technology, Soochow University, Suzhou 215006, China}
\author{Hua Jiang}
\affiliation{College of Physics, Optoelectronics and Energy, Soochow University, Suzhou 215006, China}
\author{X. C. Xie}
%\email{xcxie@pku.edu.cn}
\affiliation{International Center for Quantum Materials and School of Physics, Peking University, Beijing 100871, China}
\affiliation{ Collaborative Innovation Center of Quantum Matter, Beijing, China}
 	
 	\date{\today}
 	
 	\begin{abstract}
 	Scaling theory predicts complete localization in $d=2$ in quantum systems belonging to orthogonal class  (i.e. with time-reversal symmetry and spin-rotation symmetry).  The conductance $g$ behaves as $g \sim exp(-L/l)$ with system size $L$ and localization length $l$ in the strong disorder limit. However, classical systems can always have metallic states in which Ohm's law shows a constant $g$ in $d=2$.  We study a two-dimensional quantum percolation model by controlling dephasing effects. The numerical investigation of $g$ aims at simulating a quantum-to-classical percolation evolution. An unexpected metallic phase, where $g$ increases with $L$, generates immense interest  before the system becomes completely classical. Furthermore, the analysis of the scaling plot of $g$ indicates a metal-insulator crossover.
 	\end{abstract}
 	
 	
 	\maketitle
 	
 	\section{Introduction}
 	 It has been understood that the scaling properties of the conductance $g$  are determined by one-parameter scaling theory \cite{Anderson1979,Altland1997}. The scaling function, namely $\beta$ function reads $\beta(g)=d(\ln g)/d(ln L)$\cite{Anderson1979}, where $g$ is the conductance and $L$ is the size of the sample. When $\beta=0$, the system is at the transition point. Positive $\beta$ shows that the conductance $g$ increases with the system size $L$ indicating a metallic state. The conductance $g$ decreases with the size $L$ when $\beta< 0$ characterising an insulating state. The $\beta$ function is always negative in $d=2$ for a quantum system belonging to orthogonal class. Thus, there is no metal-insulator transition(MIT) in $d=2$ for arbitrary weak disorder, according to the scaling theory.  The conductance $g$ behaves as $g\sim exp(-L/l)$ with system size $L$ and localization length $l$. However, it is well known that there could be a MIT in the classical systems. Ohm's law tells us that the conductance $g$ is constant for a classical metal regardless of $L$ in $d=2$. The performances of $g$ in the quantum and classical systems are very different. Phase coherence is one of the key features that determines whether a system is quantum or classical. However, phase coherence can be easily lost in real systems, and the system tends to be a classical one under dephasing effects. Thus, a metallic state may appear instead of complete localization.  The mechanism of how the phase coherence affects the MIT in a quantum-to-classical evolution is not clear until now.

   	\begin{figure}[ht]%"[]"中为位置参数,四个参数tbph 依次是置顶〠ç½®åº•ã€æµ®åŠ¨ã€ å½"前位置,,选用的参数优先顺序为h-t-b-p
 		\centering
 		\includegraphics[scale=0.45]{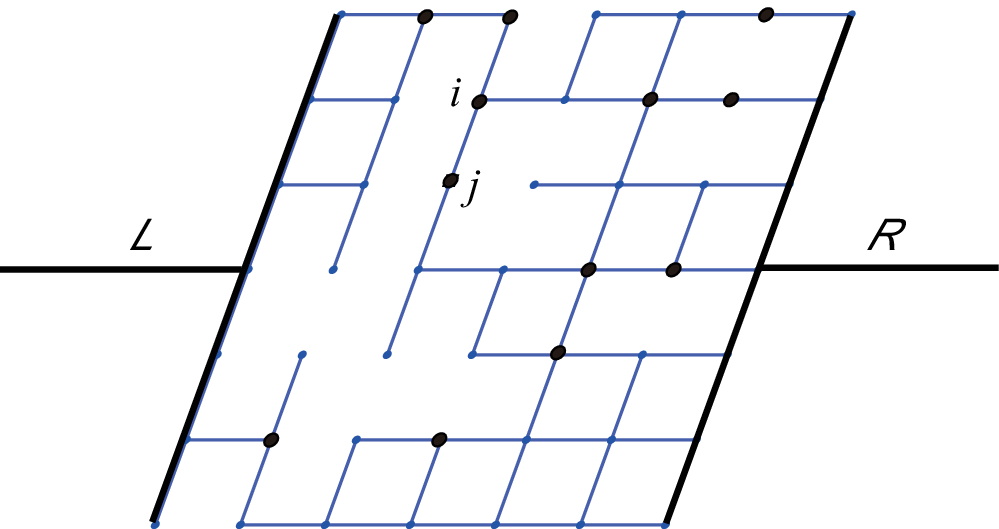}%"scale" 后的数字为图形的宽度,也可用"width=1.0\columnwidth" 定义
 		\caption{(Color online) A percolation lattice model with size $N=L\times L$ is sandwiched between the left(L) and right(R) leads. The sites with  black dots $\bullet$ are connected to the  B{\"u}ttiker's virtual leads randomly.} % 图题
 		\label{fig:figure1}
 	\end{figure}
 	
 	 In the present paper, we study a two-dimensional(2D) quantum percolation model\cite{Shapir1982,Meir1995,Chang1995,SJR1999} describing the dynamics of a quantum particle moving in a random system. According to one-parameter scaling theory, all the states in such a 2D quantum system are localized under the Anderson disorder\cite{MacKinnon,Evers2008}. On the other side, classical percolation theory always has a threshold $P_c$ of the percolation transition. Therefore, we investigate the evolution process in details. We introduce the dephasing mechanism  to destroy the quantum coherence for the purpose of switching from a quantum percolation model to a classical one. Here, we consider a two-terminal device with a central region of a 2D quantum percolation model and two ideal leads. We control in such a way that the dephasing process only takes place in central region. Specifically, the dephasing process is introduced by using B{\"u}ttiker's virtual probes\cite{Buttiker,Datta}.
 These virtual probes are coupled to the lattice sites with the current-conserving condition. We calculate the conductance $g$ for a finite-size system numerically by using the Landauer-B{\"u}ttiker formula
 combined with the non-equilibrium Green function method\cite{Landauer,Fisher1981,Meir1992,Jauho1994,Datta}. We find that an unexpected metallic phase appears before the system is entirely switched into a classical one. The conductance $g$ in the novel metallic phase increases with the system size $L$. This is not consistent with the classical Ohm's law, which shows the conductance $g$ is constant in 2D. Thus, it is more appropriate to say that the novel metallic phase contains semi-quantum and semi-classical contribution.
  %Then, we calculate the coherent length $L_\varphi$ with the disorder and dephasing strength altered.
 Furthermore, the scaling function of $g$ under different coherent lengths $L_\varphi$ is inspected  to gain deep insights. We show that the novel metallic phase maybe a consequence of metal-insulator crossover.

 	The rest of the paper is organized as follows. In Sec. II, we introduce the quantum percolation model. Dephasing mechanism is brought in to simulate a quantum-to-classical evolution. In Sec. III, we calculated the conductance numerically with the Landauer-B{\"u}ttiker formula accompanied by the non-equilibrium Green function method. We show the numerical results of the conductance with the disorder and dephasing.
 An unexpected metallic phase appears near the region $P=1$. The novel phase may contain semi-quantum and semi-classical contribution. We then investigate the scaling behaviour of the conductance $g$. The evidence shows a metal-insulator crossover. Finally, we give a brief summary in Sec. IV.
 	
 	\section{model and method}
 	
 	We start from considering a two-terminal device as shown in Fig. 1. The Hamiltonian of the central region is the 2D quantum percolation model\cite{Shapir1982,Meir1995,Chang1995,SJR1999} which can be written as

   \begin{equation}
   H=\displaystyle{\sum_{i}}\varepsilon_{i}c^{\dag}_{i}c_{i}+\displaystyle{\sum_{\left\langle i,j\right\rangle }}t_{ij}(c^{\dag}_{i}c_{j}+c^{\dag}_{j}c_{i})
   \end{equation}

  \noindent where the sum $\left\langle ij\right\rangle $ goes over the nearest neighbor sites. The on-site energy $\varepsilon_i$ obeys the uniform distribution over the interval $[-W/2,W/2]$ with the disorder strength $W$.  The bond between the nearest sites is either present $t_{ij}=1$ with probability $P$ or absent $t_{ij}=0$ with probability $1-P$.
   As shown in Fig. 1, the central region is sandwiched between the left(L) and right(R) leads with size $N=L\times L$. Unlike the central region, we assume that the L(R) leads are both ideal conductors.
  When the probability $P < P_c$ in such a percolation model, the electrons injected from the left lead cannot flow through the central region to the right lead. Thus, there is no current in the device. Once the probability excess the threshold value $P_c$, current can flow into the right lead. The conductance is calculated by applying the Landauer-B{\"u}ttiker formula
 combined with the non-equilibrium Green function method\cite{Landauer,Fisher1981,Meir1992,Jauho1994,Datta}. We only add
 the dephasing effects in the central region. The dephasing mechanism is introduced
 by using B{\"u}ttiker's virtual probes\cite{Buttiker,Datta} to simulate the quantum to classical percolation evolution. We assume  that the lattice sites are randomly chosen to be connected to the virtual leads with the dephasing probability $p_v$ and the dephasing strength $t_v$. The black dots in Fig. 1 shows  the lattice sites $i$ which  are coupled by the virtual leads. There are totally $N_v=p_v\times N$ virtual leads in the central region.

We add a small bias $V=V_L-V_R$ between the left lead and right lead, which can drive a current flowing along the longitudinal direction. Either real or virtual lead current $I_p(p=L,R,1,2,\ldots,N_v)$  is given by multiprobe Landauer-B{\"u}ttiker formula\cite{Buttiker,Datta}

     \begin{equation}
I_p=\frac{2e}{h}\displaystyle{\sum_{q\neq p}} T_{p\leftarrow q}(V_p-V_q),
  \end{equation}

  \noindent where $V_p$ is the bias in the lead $p$. The transmission function from lead  $q$ to lead $p$ is expressed as $T_{p\leftarrow q}=$Tr$[\Gamma_p G^{r} \Gamma_q G^{a}]$, where the line width function  $\Gamma_{p}=i(\Sigma_{p}^{r}-\Sigma_{p}^{r \dagger})$, with the retarded self-energy  $\Sigma_{p}^{r}$.
  The retarded Green function can be calculated by $G^{r}=[G^{a}]^{\dagger}=[EI-H-\sum_{p}\Sigma_{p}^{r}]^{-1}$, where $E$ is the Fermi energy. After we get the current $I_L$, the conductance can be directly obtained as $g=(V_L-V_R)/I_L$. We note that the percolation probability $P$, the dephasing probability $p_v$ and the dephasing strength $t_v$ can affect the coherent length $L_{\phi}$ remarkably\cite{XYX}.

  	\begin{figure}[ht]
 	\centering
 	\includegraphics[scale=0.30,bb = 5 3 1350 1550 , clip=true]{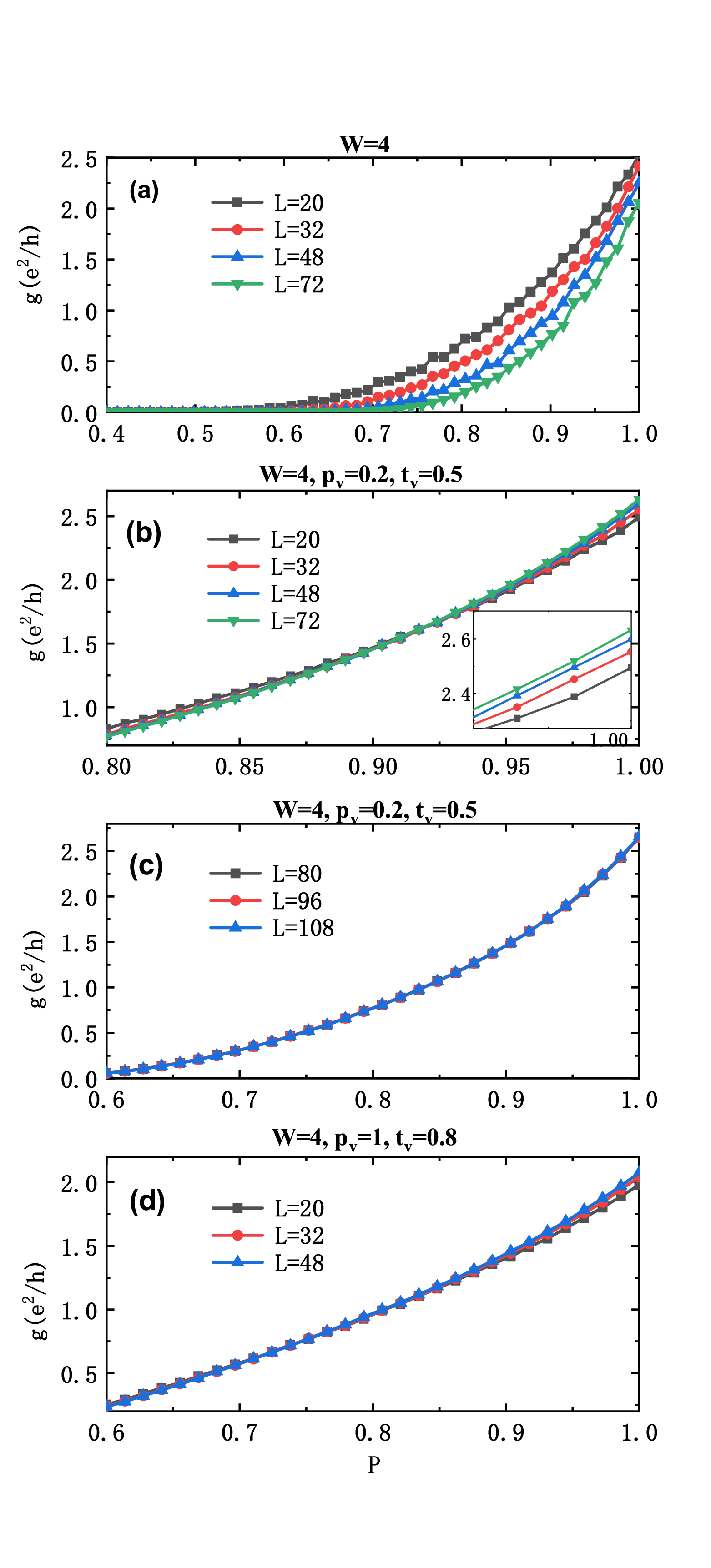}
 	\caption{(Color online) The conductance $g$ vs the probability $P$ by (a) increasing the width $L=$20, 32, 48, 72. (b) Add dephasing effects by B{\"u}ttiker's virtual probes to (a) with $p_v=0.2,t_v=0.5$. The inset is the enlargement of the metallic region.	(c) The conductance $g$ vs the probability $P$ by the width expansion of $L=$80, 96, 108. (d) Larger dephasing values are act on the model. In all subplots, we take the disorder strength $W=4$ and the Fermi energy $E=-1$ eV.	} % 图题
 	\label{fig:figure2}
 \end{figure}
 	
 	\section{unexpected metallic phase}	

 We come to the main results of our work. Firstly, let us investigate the conductance $g$ of quantum percolation model versus the probability $P$ by varying the parameters: the width $L$, the dephasing effects $p_v$ and $t_v$ at a given Fermi energy $E=-1$ eV and the disorder strength $W=4$, as shown in Fig. 2.  In quantum percolation model, the conductance $g$ is zero at small probability $P$. When $P$ increases,  electron clusters span from the one side of the lattice to the opposite side, then the conductance $g$ becomes nonzero.

 \begin{figure}[htbp]
 	\centering
 	\includegraphics[scale=0.35,bb = 5 2 650 550 , clip=true]{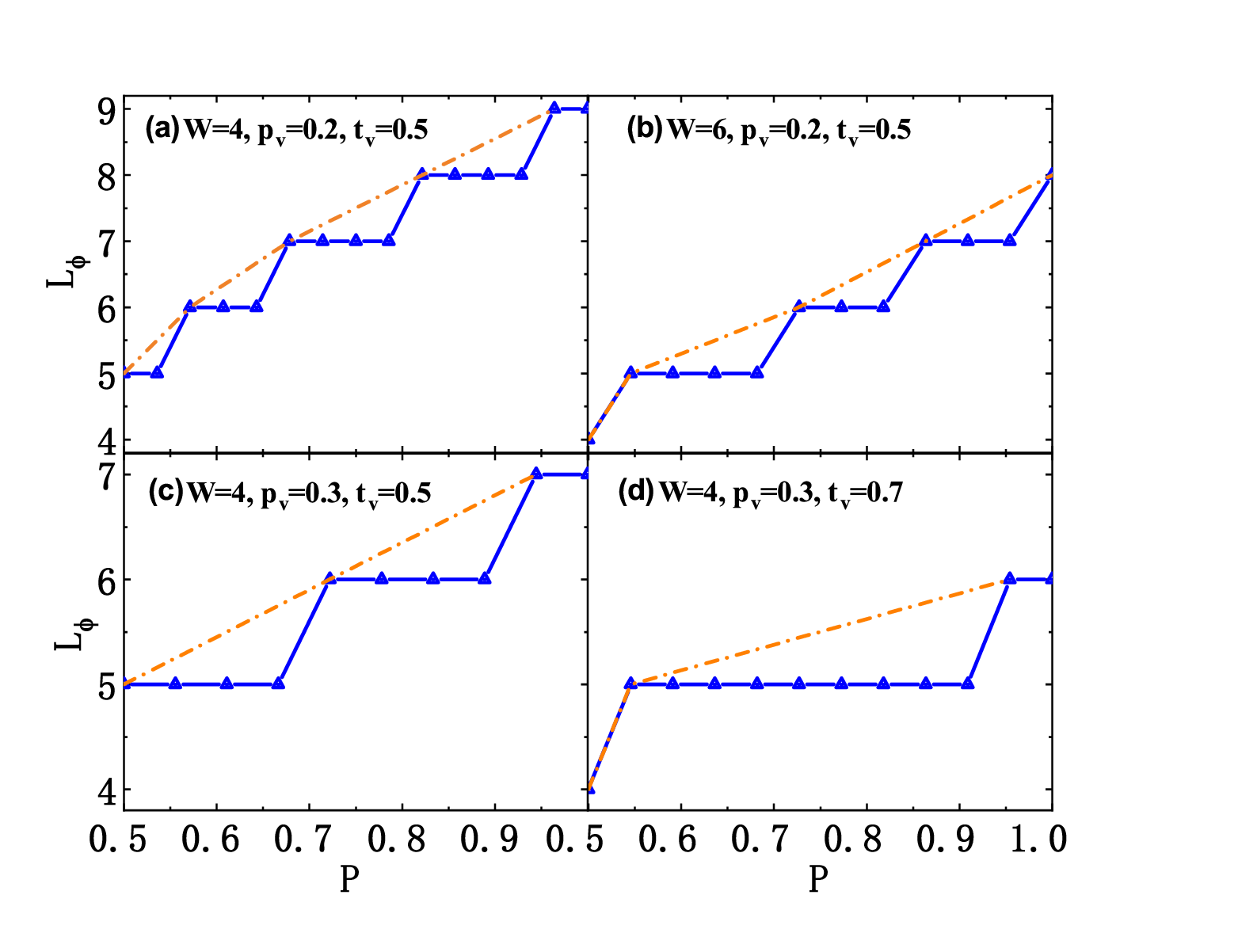}
 	\caption{(Color online) Plot the coherent length $L_{\varphi}$ vs the probability $P$ for (a) the dephasing probability $p_v=0.2$ and strength $t_v=0.5$;(c) the dephasing probability $p_v=0.3$ and strength $t_v=0.5$; (d) the dephasing probability $p_v=0.3$ and strength $t_v=0.7$. In (a),(c) and (d), the disorder strength is $W=4$. (b) We take the disorder strength $W=6$, the dephasing probability $p_v=0.2$ and strength $t_v=0.5$. The Fermi energy is at $E=-1$ eV in all cases. The dashdotted orange lines may be the possible continuous tracks of $L_\varphi$. The broken blue lines show the numerical results of $L_\varphi$. } % 图题
 	\label{fig:figure3}
 \end{figure}

  It is clearly found that the decreasing behavior of $g$ with increasing $L$ of Fig. 2(a) indicating that all the states are localized due to the disorder strength $W$. This is consistent with the one-parameter scaling theory.  Then, we bring the dephasing mechanism in the system. For example, we add $20\%$  virtual leads of the lattice sites to the system with dephasing strength $t_v=0.5$ along with the disorder. We find that all the curves of $g$ with different $L$ cross at a single point about $P_c\simeq 0.92$ (see Fig. 2(b)). More remarkably, the conductance $g$ increases monotonously with the width $L$ in the region for $P_c< P< 1$. That means an unexpected metallic phase occurs beyond the cross point $P_c$. The inset in Fig. 2(b) is the enlargement of the unexpected metallic phase. The most fascinating of the metallic phase is that its nature cannot be attributed to either the quantum class or the classical class alone. Next let us see the details. Based on the one parameter scaling theory, all the states are localized in quantum systems. Notice that the dephasing mechanism is introduced here, our model undergoes a quantum-to-classical evolution. When the dephasing strength is large enough, our system will be totally classical (see Fig. 2(d)). Although metal is common in 2D classical model, Ohm's law gives a fixed conductance $g$ regardless of the system size $L$. Thus, the novel metallic phase is not a classical one. At a moderate dephasing strength of Fig. 2(b), the system can be in a transitional situation interplay between the quantum and classical percolations. Hence, the metallic percolation phase should  contain the semi-quantum and semi-classical contributions. Furthermore, we find in Fig. 2(c) that the curves do not cross but merge when increasing the system size up to $L=108$ with the dephasing strength and disorder unchanged. This is the hallmark of Ohm's law in the classical category. As above, when we increase the dephasing strength to a larger value in Fig. 2(d), there is also an overlap trend of these curves. Consequently, we can explain the results in Fig. 2(c) and (d) from a classical perspective.

\begin{figure}[ht]
 	\centering
 	\includegraphics[scale=0.3]{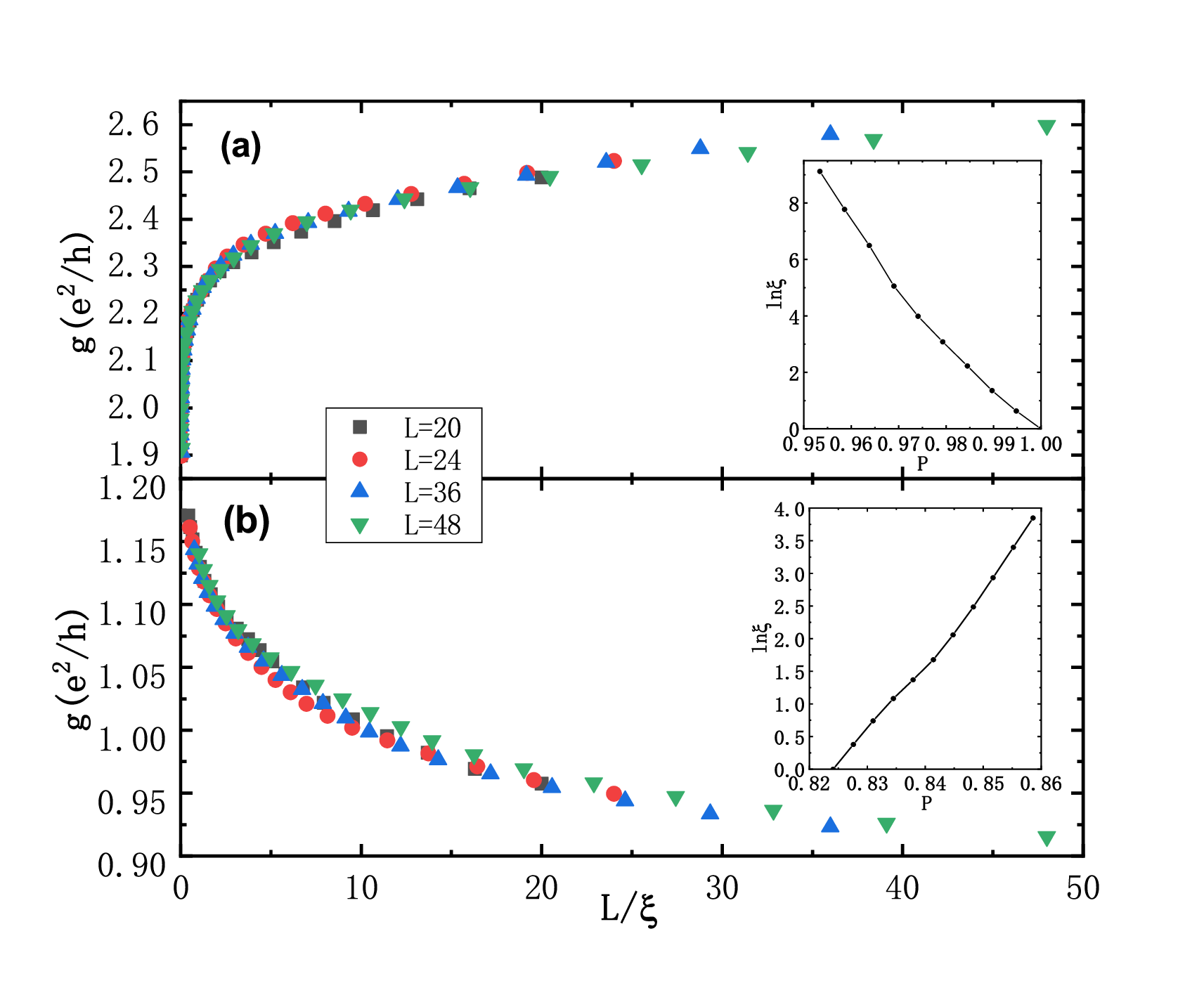}
 	\caption{(Color online) Scaling plot of $g$ vs $L/\xi$ for the system size $L=$20, 24, 36, 48 with dephasing strength $p_v=0.2,t_v=0.5$ and disorder strength $W=4$ (a) in the metallic region excess the cross point of the curves. The inset is the plot of ln$\xi$ vs the probability $P$ and (b) in the insulating region below the cross point of the curves. The inset is the plot of ln$\xi$ vs the probability $P$. 	} % 图题
 	\label{fig:figure4}
 \end{figure}

 One of the hallmarks of quantum systems, compared to classical ones, is the existence of phase coherence . Because the phase coherent length is an important length scale in quantum transport,  we will study it in details below. The current under dephasing effects contains the phase-coherent part and phase-incoherent part. When increasing the dephasing effects, the ratio of the phase-incoherent part also increases. At a certain value $p_v$ and $t_v$, both parts could have equal percentage. Thus, the system size $L$ is recognized to be equal to the phase coherent length $L_\varphi$\cite{XYX}.
 In Fig. 3, we show the phase coherent length $L_\varphi$ versus the probability $P$ at the fixed disorder and dephasing strength. It should be noted that $L_\varphi$ is a continuous value in realistic systems. The dashdotted lines may be the possible tracks of $L_\varphi$. The step-like performance of $L_\varphi$ shown in Fig. 3 is due to the constraints of numerical algorithm in which $L_\varphi$ is an integer equal to the system size $L$.
  If we add more virtual probes coupling with the lattice sites by increasing $p_v$, we find that the coherent length $L_\varphi$ gets conceivably smaller by comparing the two figures in Fig. 3(a) and (c). This agrees with common belief that dephasing always destroys the quantum coherence. In Fig. 3(d), we keep increasing the dephasing strength $t_v$  on the basis of Fig. 3(c). As a result, the coherent length $L_\varphi$ continues to decrease. In addition, the disorder strength increases, the coherent length $L_\varphi$ also decreases (see Fig. 3(b)). Now we can use the phase coherent length $L_\varphi$ obtained in Fig. 3 to uncover the physical origin of Fig. 2. In the quantum percolation limit,
 $L_\varphi \gg l$ (the localization length), the  system is strong localized with $g\sim exp(-L/l)$. Thus, the system is an insulator (see Fig. 2(a)). When the dephasing mechanism is brought in, the phase coherent length becomes finite. During the quantum-to-classical evolution, there would be a transition at a suitable $L_\varphi$. At a certain dephasing strength, $L_\varphi$ is comparable with the system size $L$ before becoming complete classical. Meanwhile, when $L_\varphi < l$ (the localization length), the system is in the ballistic-like transport, and hence a metallic behaviour appears. For a percolation system, the larger probability $P$ is, the longer the localization length $l$ is.  Thus, $l$ can exceed $L_\varphi$ near $P=1$ and is shorter than $L_\varphi$
in the region of small $P$. In particular, the region near $P=1$ presents better ballistic-like property.
 When $ L_\varphi$ keeps on getting smaller, a real sample can be divided into several phase-coherent blocks with length $L_\varphi$. In each phase-coherent  block, quantum principle is valid.  The whole sample can be viewed as an ensemble of small phase-coherent block in the classical regime. More virtual leads ($p_v$) and larger dephasing strength ($t_v$) contribute to a smaller coherent length $L_\varphi$. A similar situation occurs when increasing the system size $L$ in Fig. 2(c). In such case, the system should be classical and the conductance $g$ obeys an ohmic scaling law. When $L_\varphi \ll L$, the curves of the conductance $g$ should merge.

% \section{scaling behaviour}
At last, we follow the standard one-parameter scaling analysis\cite{MacKinnon} of  the data in Fig. 2 and show the results  in Fig. 4. The characteristic length $\xi$ is obtained by collapsing data of the conductance $g$ into a single curve. The curve represents the scaling function. We inspect the scaling behaviour of the conductance $g$ on two branches: the metallic side $P>P_c$ (see Fig. 4(a)) and the insulating side $P< P_c$ (see Fig. 4(b)). The figure shows that all the datas of conductance can merge into a single curve for different system sizes $L$. The inset is the plot of ln$\xi$ versus the probability $P$, where $\xi(P)$ diverges at $P_c$. By requiring $\xi(P) \propto \left|P-P_c \right |^{-\nu}$ in the vicinity of $P_c$, we can extract the critical exponent $\nu$.

\begin{figure}[htbp]
 	\centering
 	\includegraphics[scale=0.35]{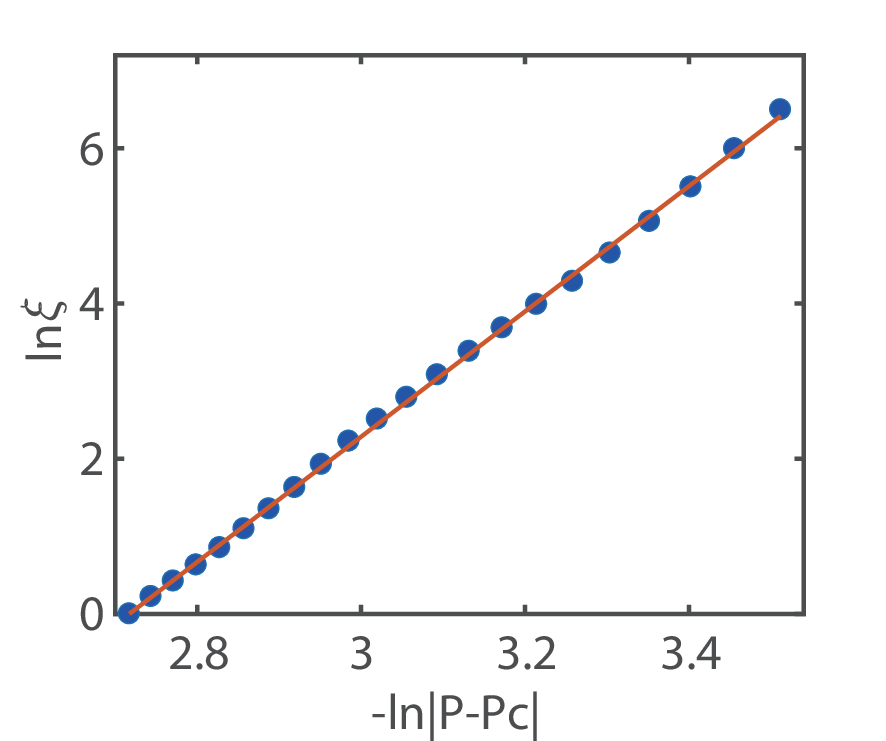}
 	\caption{(Color online) The scaling plot ln$\xi$ vs ln$ \left|P-P_c \right|$ for $P_c = 0.93$ for $L=$20, 24, 36, 48.	}
 	\label{fig:figure5}
 \end{figure}

We analyze the divergence of $\xi$ in terms of a power law $\xi(P) \propto \left|P-P_c \right |^{-\nu}$ on the metallic side shown in Fig. 5. However, we find a linear fit with slope $\nu=8.07$, much higher than any known 2D disordered systems. Based on percolation theory, the critical exponent with classical systems gives $\nu=4/3$ in d=2\cite{Stauffer1994}. Beyond this, we have inspected more scaling curves by changing the dephasing parameters. By employing a power law, the critical exponent $\nu$ is still very large with the dephasing altered.  The reason is probably that a power law is not suitable in our case.  Finally, we argue that this is due to the nature of a metal-insulator crossover. 

\section{conclusion}
 	
In conclusion, we investigate the whole evolution process from 2D quantum to classical percolation. Without dephasing, the system is a quantum percolation model with localized states.
 When increasing the dephasing effects in the quantum percolation model, the system switches towards the classical one gradually. Remarkably, an unexpected metallic phase exists at a moderate dephasing strength, and the behaviour of the conductance deviates from the classical Ohm's law. The scaling behaviour of the conductance in the presence of dephasing effects suggests a metal-insulator crossover.

 	\section*{ACKNOWLEDGMENTS}

 We gratefully acknowledge the inspiration of early work and helpful discussions with Junren Shi. They find the tendency of the metal-insulator transition\cite{SJR1999} which inspires our following research present in this work. We elaborate the cause of the metallic phase by the systematic investigation. This work is financially supported by NBRPC (Grants No. 2015CB921102, No. 2017YFA0303301, and No.
2017YFA0304600) and NSFC (Grants No. 11504008,  No. 11574245, No. 11674028 and No.
11822407).

 \end{document}